\documentclass[review,12pt]{elsarticle}
 \usepackage{graphics}
 \usepackage{epsfig}
 \usepackage{amsthm}

\usepackage{eurosym}
\usepackage{amssymb}
\usepackage{tikz}
\usepackage{amsfonts, amsmath}

\newdefinition{rmk}{Remark}
\begin{document}

\begin{frontmatter}

\title{Computational and physical consequences of interaction of closely located simultaneous hydraulic fractures}

\author[firstaddress]{Ewa Rejwer\corref{correspondingauthor}}
\cortext[correspondingauthor]{Corresponding author}
\ead{e.rejwer@gmail.com}

\author[secondaddress,thirdaddress]{Aleksandr Linkov}
 \address[firstaddress]{%
 Rzeszow University of Technology, \\
 Al. Powstancow Warszawy 12,
 35-959 Rzeszow, Poland}
 \address[secondaddress]{%
 Saint Petersburg State Polytechnical University,\\
 Polytechnicheskaya st., 29,
 St.Petersburg 195251, Russia}
  \address[thirdaddress]{%
Institute for Problems in Mechanical Engineering,
Russian Academy of Sciences,\\
V.O., Bol'shoy Pr., 61, St.Petersburg 199178, Russia}

\begin{abstract}
Strong interaction of closely located, nearly parallel hydraulic fractures and its influence on their propagation are studied. Both computational and physical aspects of the problem are considered. It is shown that from the computational point of view, when a distance between cracks is small as compared with their sizes, the system becomes ill-conditioned and numerical results deteriorate. The physical consequence of the interaction consists in decreasing of the crack opening and even greater decrease of conductivity. Then the resistance to fluid flow grows what results in the propagation of only those fractures,  the distance between which is large enough. 

The research aims to suggests a means to overcome the computational difficulty and to improve numerical simulation of hydraulic fractures in shales. Numerical experiments are carried out for a 2D problem by using the complex variable hypersingular boundary element method of higher order accuracy. The condition number of the main matrix of a system, the opening of cracks, the stress intensity factors, effective conductivity and resistance of a cluster are calculated and analysed. The results imply that to avoid computational instability, it is possible to replace a cluster of cracks by a single properly located crack. The results also evidently confirm that the propagation of distant individual fractures rather than of an array of closely located cracks is to be expected in practice, what agrees with observations. 

\end{abstract}


\end{frontmatter}

\section{Introduction}

\label{introduction} 
Hydraulic fracturing, when used to recover gas from low permeable shale structures, stimulates the flow of a fluid by generating the opening of a large number of micro cracks. Their growth and interaction with hydraulic fractures and pre-existing natural fractures affect the hydrofracture propagation and the efficiency of hydrofracturing treatment. The interaction has been studied by many authors, e.g. N.R. Warpinski  and L.W. Teufel \cite{Warpinski1987}, V.F. Koshelev and A. Ghassemi \cite{Koshelev2003} (these authors also analysed in-situ stresses along the natural fractures); X. Weng et al. \cite{Weng2011}, who attempted to model propagation of hydraulic fractures influenced by multiple natural fractures. However, because of complexity of the problem, there is still a need to understand arising difficulties and to develop means to overcome them. 

It is well-known that when the ratio distance-to-crack size decreases, the interaction becomes stronger. When cracks are nearly parallel and the ratio is small, the opening of a fracture induces compressive stresses normal to the fracture surface. This decreases the opening of neighboring, closely located fractures. As a consequence, the propagation speed of some of the fractures drops and some of them even stop. This phenomenon, called the stress shadowing, e.g. \cite{Fisher2004}--\cite{Kresse2013}, strongly affects the productivity of multiple fractures propagating from a wellbore. Thus, a rigorous analysis of the shadowing effect, tending to optimize the distance between successive treatments, is of importance for the petroleum industry. It is to be taken into account for proper modeling hydraulic fracturing.  

The paper aims to study interaction of closely located, parallel cracks under static conditions. We focus on the consequences of the interaction. In Section 2, the methods used for modeling 2D and 3D hydraulic fractures, are briefly reviewed. Section 3 contains the study of the interaction of multiple parallel cracks, symmetrically or non-symmetrically located at various distances. At the beginning, some details of the method used are presented. Numerical experiments are carried out by employing the complex variable (CV) hypersingular boundary element method, based on the CV hypersingular boundary integral equations, tailored to account for the asymptotic behavior of fields in a close vicinity of singular points, such as tips of cracks, multi-wedge points, etc. (see e.g. \cite{Linkov1994}--\cite{Rejwer2014}). For the main matrix of a system, the condition number, which is a measure of nearness to computational instability, is calculated. It is shown, that when the distance between cracks decreases, or a number of cracks in a considered area grows, the condition number grows exponentially. This indicates ill-conditioning of the main matrix, which finally causes deterioration of numerical results. Further subsections contain the study of the stress shadowing effect, the openings of cracks in a cluster and the values of the stress intensity factor (SIF) at the tips of cracks, for various distances between the cracks. Based on the calculated values of the openings, the effective conductivity of a cluster of cracks is analysed. The conclusions on the growth of the effective resistance to the flow of a fracturing fluid are made. Employing the obtained data, we conclude on the possibility to replace a cluster of cracks by a properly located single crack. This serves to decrease the condition number of the matrix to the level excluding deterioration of numerical results. The numerical results provide also conclusions on the expected propagation of individual fractures. It is stated that the results of numerical experiments are consistent with the field observations. Section 4 contains the summary of the results obtained.

\section{Brief review of the methods commonly used for modeling hydrofractures}

In this section, 2D and 3D fracture models developed for understanding hydraulic fracture process are briefly reviewed. Among them, those concerning with the analysis of the stress shadowing effect, are of special interest to our theme.

\textbf{\textit{2D fracture models and methods}}. The classical 2D models are the KGD model by S.A. Khristianovitch, Y.P. Zheltov \cite{Khristianovitch1955}, J. Geertsma and F. de Klerk \cite{Geertsma1969} and the PKN model by T. Perkins, L. Kern \cite{Perkins1961} and R. Nordgren \cite{Nordgren1972}. The KGD model assumes the plane strain state in planes perpendicular to the fracture front, 
and it is used to study fractures on the initial stage of the propagation. In the PKN model, the plane-strain state is assumed in planes parallel to the fracture front. 
The both problems refer to a \textit{homogeneous} medium and they are reduced to solving in time a system of 1D nonlinear differential (for the PKN model) or integro-differential (for the KGD model) equations in a single spatial coordinate. Presently the models account for a pumping regime and Carter's leak-off.


An approach for modeling a 2D fracture in an \textit{inhomogeneous} medium with a net of cracks employs the distinct element method (DEM), proposed by P.A. Cundall \cite{Cundall1979} for solving rock mechanics problems. The DEM can simulate rocks subjected to static or dynamic loading. A region is divided by a discrete fracture network (DFN) that separates a finite number of discrete blocks. Fluid leak-off is accounted for automatically as fluxes in cracks branching from the major fractures. 

\textbf{\textit{3D fracture models}}. 3D models serve to study fracture nucleation and growth in a 3D space. Some of them  account for multi-layered structure of rock formation. Specifically, the pseudo 3D model (P3D, e.g. \cite{Mack2000}) modifies the PKN model, adding height variation along the fracture length. Later on, the P3D model was adjusted to studying tight-gas formations in the form of so-called unconventional fracture model (UFM, \cite{Weng2011}, \cite{Kresse2013}). The UFM serves to simulate fracture propagation, rock deformation, and fluid flow in the complex fracture network created during a hydrofracturing treatment. It combines a number of parallel and orthogonal PKN fractures. To account for the stress shadowing effect, it is supplied with corrective coefficients found from particular benchmark solutions. This also serves to roughly simulate the interaction of hydraulic fractures with pre-existing natural fractures. The model requires calibration of its parameters.   

An alternative way to overcome the limitations of P3D model consists in employing 3D variants of the DEM and DFN (\cite{Damjanac2010}--\cite{Nagel2011}, \cite{Cundall1985}, \cite{Pettitt2011}). These models combine geomechanical models and field observations. They are developed in mining and civil engineering and they are also used for coupled processes, including stimulation of fractured rock mass by fluid injection, propagation of hydraulic fracture and its interaction with DFN. Still the DEM and DFN models are computationally expensive and they require calibration of parameters characterizing elastic, fluid and fracture properties. The calibration is performed by using available benchmark solutions of appropriate problems. 

The models discussed have been successfully employed for modeling fracture geometry and/or interactions between hydrofractures and pre-existing natural fractures. However, they do not allow to distinctly evaluate mutual influence of nearly parallel cracks, when the distance between them becomes notably less than the crack in-plane sizes. In view of significance of such configurations, we focus on their detailed investigation.

\section{Interaction of closely located fractures}	

Strong interaction of nearly parallel closely located fractures concerns with both the computational and physical aspects of the problem. On one hand, there appear boundaries at small distances between them, so that after discretization the nodal points of meshes become very close; in limit, the boundaries coincide. Hence any numerical method, not specially designed for accounting for the asymptotic behavior of a solution in a thin strip between the boundary surfaces, unavoidably deteriorates when the thickness of the strip is too small. The situation is similar to that considered in the theories of thin plates and shells. Direct solution of exact equations becomes impossible with decreasing thickness of a shell. Meanwhile the asymptotic analysis and proper using of the specific geometry provide radical simplifications and very accurate solutions of important engineering problems. Actually the strip between nearly parallel closely located cracks is in the state similar to that in a plate of the same thickness. Consequently it may be expected that some features of the plate deformation and bending will be reproduced in the strip between cracks. Meanwhile the degree of similarity and the edge effects are uncertain, and they are to be thoroughly examined by using a proper numerical method. The later should provide accurate and stable results up to the thickness, for which the asymptotic behavior of a solution becomes evident. 

Note that the thickness, for which asymptotics start to dominate, is not known in advance. This impels using of (i) a method of high accuracy and (ii) stability control of numerical results obtained. Below we meet these two requirements and conclude on possible simplifications accounting for the asymptotic behavior of the solution.   
 
On the other hand, there arise physical effects specific for the hydrofracture problem. In particular, the shadowing effect, mentioned in many papers on hydraulic fracture simulation, manifests itself in drastic difference of the stresses in the strip between nearly parallel closely located cracks as compared with stresses near the surface of a single crack. It is just one of the physical effects, generated by the crack interaction. It is closely related to other effects, some of which are even of greater significance for the fracture propagation. The analysis below tends to reveal and to quantify them.

\subsection{Brief description of the method used for calculations}

	The calculations are carried out by using the complex variable hypersingular boundary element method (CV H-BEM) for the equation \cite{Dobroskok2009}
\begin{equation}\label{equ2_21}
Re\left(-\frac{e^{i\alpha_z}}{2\pi}\int^{}_{L}\left[ie^{-i\alpha_\tau}\frac{\Delta (q_n/k)}{\tau -z}+\frac{\Delta \phi}{(\tau -z)^2}\right]d\tau\right)=0.5\left(\frac{q_n^+}{k^+}+\frac{q_n^-}{k^-}\right),~~z\in L
\end{equation}\smallskip
where $L$ is the total boundary of all cracks in a cluster, in which thermal and/or mechanical values may experience discontinuity; $z\in L$ is the complex coordinate of a field point; $\tau$ is the complex coordinate of an integration point; the symbol $\Delta$ in front of a value, denotes its jump across the contour; $\varphi$ is the potential (displacements); $q_n$ is the normal component of the flux (traction vector) at an element of $L$ with the normal $n$; $k=-k_f$, $k_f$ is the conductivity (shear modulus) of a matrix; $\alpha_z$ ($\alpha_{\tau}$) is the angle between an element $dz$ ($d\tau$) and the $x$-axis; the index "plus" ("minus") refers to the side, with respect to which the normal $n$ is outward (inward). 

The CV H-BIE (\ref{equ2_21}) refers to anti-plane deformation and consequently $\Delta\phi$ is the discontinuity of the shear displacement through a crack. Still the equation  keeps the general features of crack interaction and it serves us to simplify the analysis before studying more computationally involved plane-strain and 3D problems. (An extension to the latter problems on the basis of the results obtained is in progress). To keep track with the hydraulic fracturing we shall associate the displacement discontinuity with the opening.   
	
We solve the problem by employing the CV H-BEM of higher accuracy. To this end, higher-order approximation of the density functions and accounting for singular behavior of the solution near crack tips are used. Specifically, the approximation includes CV polynomials of an arbitrary order and the power function with an arbitrary rational exponent $\beta$:
\begin{equation}\label{equ2_28}
f_{j}(\tau)=\left( \tau-c\right) ^{\beta}\tau ^{ j},  
\end{equation}
where $c$ is the local complex coordinate of the end point of a boundary element; for a standard straight element $j=0,1,2,...$, and for a standard circular-arc element $j=0,\pm 1,\pm 2,...$; the exponent $\beta$ accounts for the asymptotics of the density function near the singular points (e.g. tips of cracks). Thus, for an ordinary element $\beta = 0$. For a singular element, $\beta= m/n\in (-1, 1)$ is a rational fracture, which in general is found by using a standard subroutine (see e.g. \cite{Linkov2006}).

For each boundary element, the global coordinates are transformed to its local coordinates so that an element becomes standard \cite{Linkov2002}. After using the approximation (\ref{equ2_28}) in the local coordinates, the CV singular and hypersingular, entering the CV H-BIE (\ref{equ2_21}), for an element are
\begin{equation}\label{equ2_29}
S_{j}(z)=\int\limits_{b}^{c}\frac{f_{j}(\tau)d\tau}{\tau -z},~~~~H_{j}(z)=\int\limits_{b}^{c}\frac{f_{j}(\tau )d\tau }{(\tau -z)^{2}}. 
\end{equation}

Without loss of generality, we assume $b$, $c$ to be start and end point of the standard boundary element, respectively. When the standard element is singular, the $c$ point at the tip of crack represents a singular point. For a standard straight element $b=-1$, $c=1$; for a circular-arc element with the angle $2\theta _{0}$, $b=exp(-i\theta _{0})$, $c=exp(i\theta _{0})$. Below we use straight elements only.

Evaluation of the influence coefficients (\ref{equ2_29}) over ordinary and singular elements is easily performed by employing the recurrent analytical formulae (see e.g \cite{Linkov2002})
\begin{equation}\label{equ2_30}
S_{j+1}=zS_{j}(z)+a_{j},~~~~H_{j+1}=zH_{j}(z)+S_{j}(z),~~~~j=0,1,2,..., 
\end{equation}
\begin{equation}\label{equ2_31}
S_{j-1}=\frac{1}{z}[S_{j}(z)-a_{j-1}],~~~~H_{j-1}=\frac{1}{z}
[H_{j}(z)-S_{j-1}(z)],~~~~j=0,-1,-2,...,  
\end{equation}
where $a_{j}=\int\limits_{b}^{c}(\tau -c)^{\beta }\tau ^{j}d\tau $ are constants evaluated analytically. It should be emphasized, that the formulae (\ref{equ2_30}), (\ref{equ2_31}) for the influence coefficients are exact and they notably decrease the time expense, as compared with numerical integration. 

The starting integrals $S_0$ and $H_0$ in (\ref{equ2_30}) and (\ref{equ2_31}) are:
\begin{equation*}
S_{0}(z)=\int\limits_{b}^{c}\frac{(\tau -c)^{\beta }d\tau }{\tau -z}=\int\limits_{b-c}^{0}\frac{\tau ^{\beta }d\tau }{\tau -t},~~~~H_{0}(z)=\int\limits_{b}^{c}\frac{(\tau -c)^{\beta }d\tau }{(\tau -z)^{2}}=\int\limits_{b-c}^{0}\frac{\tau ^{\beta }d\tau }{(\tau -t)^{2}},
\end{equation*} 
where $t=z-c$. They are also evaluated analytically. 
Specifically, for an ordinary element ($\beta=0$) they are:
\begin{equation*}
S_0(z)=\ln\frac{c-z}{b-z}+\pi i\delta_z,~~~~H_0(z)=\frac{1}{b-z}-\frac{1}{c-z},
\end{equation*}
where $\delta _{z}=1$ for $z\in\lbrack b,c]$, $\delta _{z}=0$ for $z\notin\lbrack b,c]$, $z=|z|\exp {\left(i\gamma\right)}$.

For a singular element ($\beta\neq 0$), it is sufficient to consider the case when $\beta>0$. If $\beta<0$ ($|\beta|<1$), we may write $\beta=\beta_1 -1$ with $\beta_1=\beta +1>0$. Then $\tau^\beta =\tau^{\beta_1} \tau^{-1}$ and we arrive at the integrals $S_{-1}(z)$, $H_{-1}(z)$ with positive $\beta_1$ ($0<\beta_1<1$). Analytical formulae for the starting integrals $S_{0}(z)$, $H_{0}(z)$ with positive $\beta=m/n$ ($n>m $), have been given in \cite{LinkovKB2002}. We present them in the simplified forms: 
\begin{equation}  
S_{0}(z)=\int\limits_{b-c}^{0}\frac{{\tau}^{\beta}}{\tau -t}d\tau =\frac{-1}{\beta }(b-c)^{\beta} +\sum\limits_{s=0}^{n-1}t_{s}^{m}Ln{\frac{t_{s}}{t_{s}-\sqrt[n]{b-c}}}+\pi it_{0}^{m}\delta _{t}, \label{equ8}
\end{equation}
\begin{equation}  
H_{0}(z)=\int\limits_{b-c}^{0}\frac{{\tau}^{\beta}}{\left(\tau -t\right) ^{2}}d\tau =\frac{\beta}{t}S_{0}+\frac{(b-c)^{\beta}}{b-c-t}+\frac{1}{t}(b-c)^{\beta }, \label{equ9}
\end{equation}
where $t=z-c$, $\delta _{t}=1$ for $t\in\lbrack b-c,0]$, $\delta _{t}=0$ for $t\notin\lbrack b-c,0]$, $t=|t|\exp {\left(i\eta\right) }$, $t_{s}=|t|^{\frac{1}{n}}\exp {\left(i\left(\eta +2s\pi \right)/n\right)}$ is a root of
the complex value $t$. The root $\sqrt[n]{b-c}$ is the principal one ($s=0$). 

Numerical tests and experience, gained to the date, show that these forms of the CV BEM provide accurate and stable results (see e.g. \cite{Linkov2002}, \cite{Rejwer2014}, \cite{Dobroskok2009}, \cite{Mogilevskaya1996}). The results of numerical experiments, presented in this paper, are obtained by employing the CV H-BIE (\ref{equ2_28}) with the second order approximation of the density functions (displacement discontinuities and tractions) and a square root asymptotics near the tips of cracks ($\beta =0.5$).

\subsection{Calculation of the condition number}

Note that when the number of cracks in a structure increases, the probability of configurations with almost parallel, closely located cracks is high. When the distance between parallel cracks tends to zero, the main matrix contains rows with practically the same influence coefficients. This leads to ill-conditioning of the matrix and causes significant computational difficulties. Then, as mentioned, a numerical method not specially designed to account for this geometrical feature unavoidably fails and numerical results deteriorate. Therefore, it is essential to trace the condition number (see e.g. \cite{Epperson2002}), which is a measure of nearness to computational instability. 
	
    \begin{figure}[htp]
    \centering
 \begin{minipage}[t]{0.97\linewidth}
	\begin{tikzpicture}[scale=6.9]
\draw[white, arrows=->, line width=0.4pt] (-0.2,0.3)--(-0.2,0.4);
\draw[white, arrows=->, line width=0.4pt] (-0.2,0.3)--(-0.1,0.3);
\draw(-0.16,0.4) node[text=white]{$y$};
\draw(-0.06,0.31) node[text=white]{$x$};
\draw(-0.16,0.66) node[text=black]{$a)$};

\draw(0.44,0.6) node[text=black]{\small $2\_1$};
\draw(0.44,0.45) node[text=black]{\small $2\_2$};
\draw(0.986,0.6) node[text=black]{\small $4\_1$};
\draw(0.986,0.55) node[text=black]{\small $4\_2$};
\draw(0.986,0.50) node[text=black]{\small $4\_3$};
\draw(0.986,0.45) node[text=black]{\small $4\_4$};
\draw(1.55,0.6) node[text=black]{\small $6\_1$};
\draw(1.55,0.525) node[text=black]{$...$};
\draw(1.55,0.45) node[text=black]{\small $6\_6$};

\draw[black, line width=1.3pt](0.0, 0.6 )--(0.396, 0.6);
\draw[black, line width=1.3pt](0.0, 0.45)--(0.396, 0.45);
\draw[black, line width=1.3pt] (0.55, 0.6)--( 0.946, 0.6);
\draw[black, line width=1.3pt](0.55, 0.55)--(0.946,0.55);
\draw[black, line width=1.3pt](0.55, 0.5)--( 0.946, 0.5);
\draw[black, line width=1.3pt](0.55, 0.45)--(0.946, 0.45);
\draw[black, line width=1.3pt]( 1.1, 0.6)--( 1.496, 0.6);
\draw[black, line width=1.3pt]( 1.1, 0.57)--( 1.496, 0.57);
\draw[black, line width=1.3pt](1.1, 0.54)--( 1.496, 0.54);
\draw[black, line width=1.3pt](  1.1, 0.51)--( 1.496, 0.51);
\draw[black, line width=1.3pt]( 1.1, 0.48)--( 1.496, 0.48);
\draw[black, line width=1.3pt]( 1.1, 0.45)--( 1.496, 0.45);

\draw[dashed, arrows=<->, line width=0.4pt] (0.0,0.39)--(0.396,0.39);
\draw(0.198,0.34) node[text=black]{\small $L=0.15$};
\draw[dashed, arrows=<->, line width=0.4pt] (-0.04,0.45)--(-0.04,0.6);
\draw(-0.08,0.525) node[text=black]{\small $d$};
	\end{tikzpicture}
	\end{minipage} 
	\begin{minipage}[t]{0.97\linewidth}
	\begin{tikzpicture}[scale=6.9]
\draw[arrows=->, line width=0.4pt] (-0.16,0.2)--(-0.16,0.3);
\draw[arrows=->, line width=0.4pt] (-0.16,0.2)--(-0.06,0.2);
\draw(-0.12,0.3) node[text=black]{$y$};
\draw(-0.02,0.21) node[text=black]{$x$};
\draw(-0.16,0.68) node[text=black]{$b)$};

	\draw(0.44,0.6) node[text=black]{\small $2\_1$};
\draw(0.31,0.45) node[text=black]{\small $2\_2$};
\draw(0.92,0.6) node[text=black]{\small $4\_1$};
\draw(0.853,0.55) node[text=black]{\small $4\_2$};
\draw(0.986,0.50) node[text=black]{\small $4\_3$};
\draw(0.92,0.45) node[text=black]{\small $4\_4$};
\draw(1.41,0.6) node[text=black]{$6\_1$};
\draw(1.54,0.525) node[text=black]{$...$};
\draw(1.48,0.45) node[text=black]{$6\_6$};

\draw[black, line width=1.3pt](0.132, 0.6)--(0.396, 0.6);
\draw[black, line width=1.3pt]( 0.0, 0.45)--(0.263, 0.45);
\draw[black, line width=1.3pt](0.615, 0.6)--(0.879, 0.6);
\draw[black, line width=1.3pt]( 0.55, 0.55)--(0.813, 0.55);
\draw[black, line width=1.3pt]( 0.681, 0.5)--( 0.946, 0.5);
\draw[black, line width=1.3pt]( 0.615, 0.45)--(0.879, 0.45);
\draw[black, line width=1.3pt]( 1.1, 0.6)--( 1.363, 0.6);
\draw[black, line width=1.3pt]( 1.231, 0.57)--( 1.496, 0.57);
\draw[black, line width=1.3pt](1.165, 0.54)--( 1.429, 0.54);
\draw[black, line width=1.3pt]( 1.1, 0.51)--( 1.363, 0.51);
\draw[black, line width=1.3pt]( 1.231, 0.48)--( 1.496, 0.48);
\draw[black, line width=1.3pt]( 1.165, 0.45)--( 1.429, 0.45);

\draw[dashed, arrows=<->, line width=0.4pt] (0.0,0.41)--(0.263,0.41);
\draw(0.1315,0.36) node[text=black]{\small $L=0.1$};
\draw[red, dashed, arrows=<->, line width=0.4pt] (0.0,0.31)--(0.396,0.31);
\draw(0.193,0.27) node[text=red]{\small $0.15$};
\draw[dashed, arrows=<->, line width=0.4pt] (-0.04,0.45)--(-0.04,0.6);
\draw(-0.08,0.525) node[text=black]{\small $d$};
	\end{tikzpicture}
	\end{minipage}
	\caption{The schemes for $N=2$, $4$ and $6$ of a) symmetrically and b) non-symmetrically located, parallel cracks of the length $L$ and the distance between external cracks $d$.}
\end{figure}
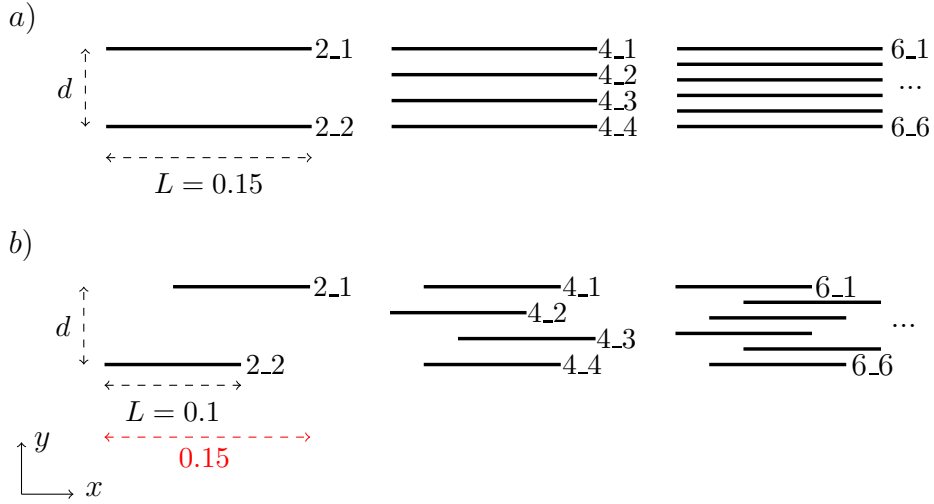

As an example, we consider the clusters of $N=2$, $3$, $4$, $6$ and $8$ straight parallel cracks of the same length $L=2l=0.15$ for symmetrical cracks, and $L=2l=0.1$ for non-symmetrical cracks, located in a square region with the side length equal to $1.0$. The square is subjected to the anti-plane shear. The distance between the external cracks in a cluster is $d$ (see Fig.1). In the numerical experiments, the distance $d$ decreases from $0.6l$ to $0.006l$. The cracks are located either symmetrically (Fig.1a) or non-symmetrically (Fig.1b) about the horizontal middle cross-section. Each crack is discretized by four, three node straight boundary elements: two ordinary elements near the center of a crack and two singular at the tips of cracks.  In calculations we assume: zero tractions at crack surfaces; zero and unit displacements at the left and right vertical sides of the square, respectively; and zero shear traction $\sigma^0_{zx}=0$ at its horizontal sides. The condition number is calculated for different numbers $N$ of cracks in a cluster and for various distances $d$.

The condition number for symmetrically located cracks exceeds $10^{10}$--$10^{12}$ when $d/L$ is less than $0.04$ (Fig.2). For non-symmetrically located cracks the interaction is weaker and the condition number grows slower (Fig.3), as compared with symmetrically located cracks. Still the condition number for them exceeds $10^8$--$10^{10}$ when the ratio $d/L$ becomes less than $0.04$.

\begin{figure}[htp]
\centering
 \includegraphics[width=0.81\textwidth]{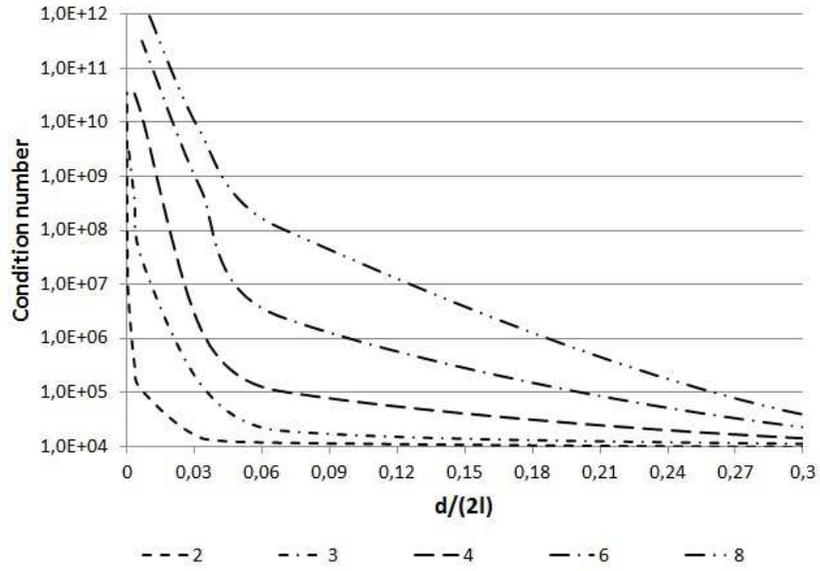}
\caption{Condition number for symmetrically located cracks.}
\end{figure}
\begin{figure}[hbp]
\centering
 \includegraphics[width=0.81\textwidth]{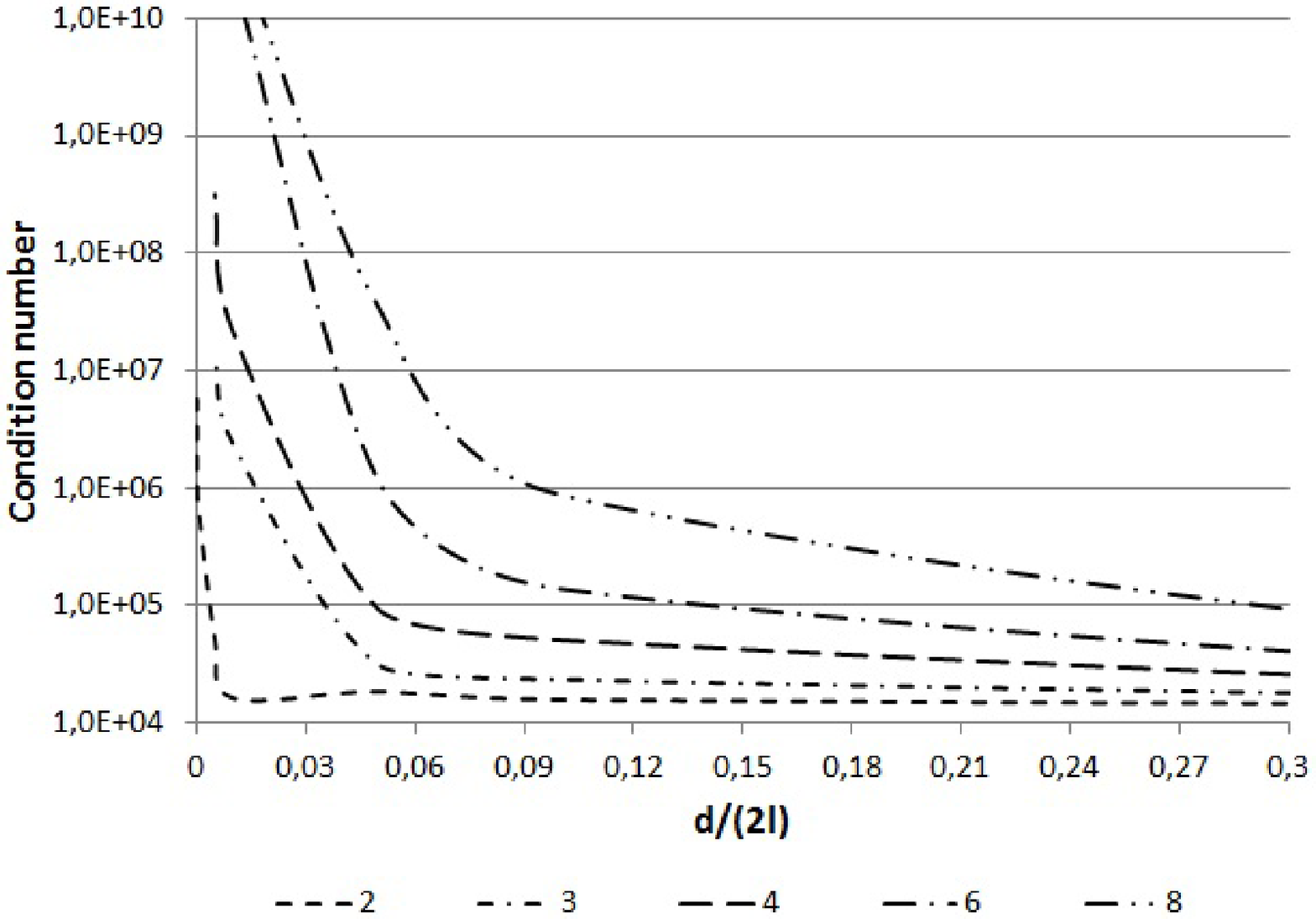}
\caption{ Condition number for non-symetrically located cracks.}
\end{figure}

It is essential, that in the both cases, when the distance $d$ decreases and/or the number $N$ of cracks in a cluster grows, the condition number grows \textit{exponentially}. Moreover, based on the results of numerical tests, we could see that the destabilizing effect of the interaction between cracks becomes significant when the condition number exceeds the values $10^7$--$10^{9}$. Then the computational error is well above the prescribed tolerance. 

These results refer to a \textit{single} cluster of cracks. When modeling strongly inhomogeneous structures (i.e. shale structures, etc.) with a high crack density, the number of clusters of closely located, nearly parallel cracks increases. According to the performed calculations, this may drastically complicate the numerical analysis of a problem.

\subsection{Analysis of stress shadowing, crack opening, stress intensity factors and conductivity}

The analysis of quantities characterizing mutual influence of the cracks in a cluster is performed for the same configurations as those considered in the previous subsection (Fig.1). Calculations show that in the strips between closely located cracks, the stresses on areas parallel to cracks surfaces are almost constant and they are the same as those on the surfaces themselves. This is the so called \textit{stress shadowing effect}. It results in almost uniform strains in the strips between the cracks so that mutual displacements of two surfaces of a strip is practically zero when the thickness of a strip is small (recall that for a uniform strain, these mutual displacements approximately equal to the product of the strain by the strip thickness). Hence it could be expected that the openings are localized only at the first and the last cracks of a cluster, while the openings of internal cracks are practically zero. The calculations of the openings confirm this suggestion.\\

\begin{figure}[hp]
\centering
  \includegraphics[width=0.83\textwidth]{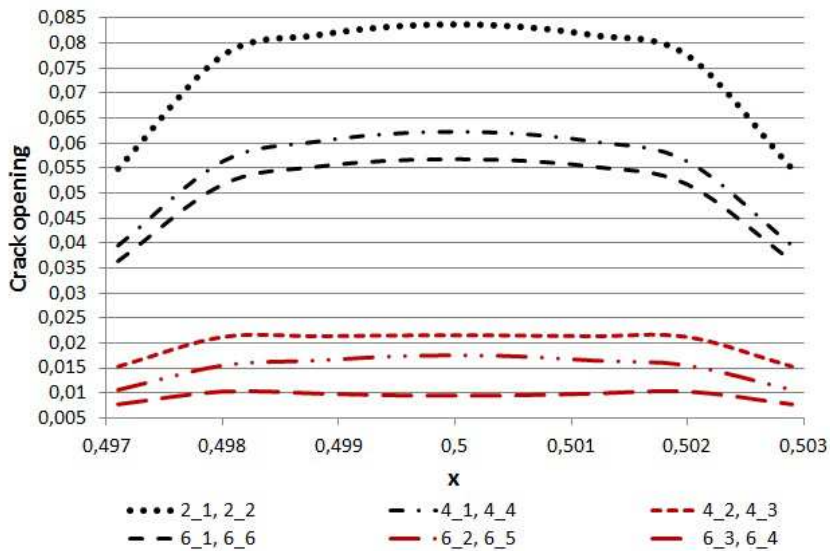}
  \caption{Crack openings along $N=2$, $4$ and $6$ symmetrically located cracks in a cluster.}
\end{figure}

Fig.4 presents the distributions of the openings along each of the cracks in a cluster of $2$, $4$ and $6$ parallel cracks. The cracks are located at the distance $d=0.01$ ($d/L=0.067$). Clearly, in accordance with the suggestion, the openings of internal cracks are much less than those of the external ones. Specifically, the openings of the first and the $N$-th crack approximately equal to $w_{C_1} = w_{C_N} \approx w_S/2$, while the openings of the intermediate cracks tend to zero. The major input into the sum is that of the external cracks. 

This effect becomes stronger with growing number of cracks in a cluster with fixed distance $d$ between the external cracks (see Fig.4). We conclude that when the distance between cracks decreases, the sum of the openings $\sum_{i=1}^Nw_{C_i}$ tends to the opening of a single crack $w_S$ (for the considered example, $w_S=0.148$ at the center of the crack).  

Note that the distances, at which the discussed asymptotic distributions of the stresses, stains and displacements in the thin strips between cracks become evident, are well beyond the range of computational instability of the method employed. The results start to deteriorate merely at the ratio $d/L=0.0003$, $0.006$ and $0.033$ for $N=2$, $4$ and $6$, respectively. Meanwhile, in the examples the considered ratio was notably greater ($d/L=0.067$). 

For comparison, the values of openings calculated for the cluster of $N=2$ and $4$ \textit{non-symmetrically} located cracks, are presented in Fig.5 and Fig.6, respectively. We see, that at points "shadowed" by other cracks the openings are less. In particular, for a cluster of $N=4$ cracks the stress shadowing, induced by the internal cracks, strongly decreases the openings of the external cracks.\\

\begin{figure}[hbp]
\centering
  \includegraphics[width=0.8\textwidth]{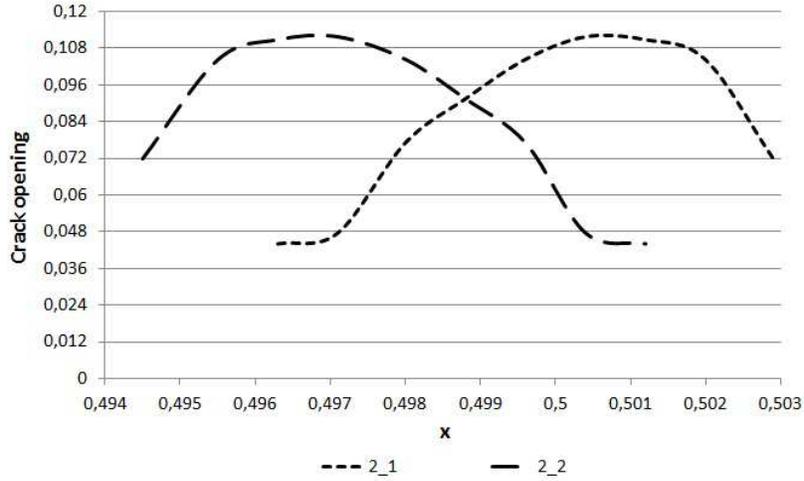}
  \caption{Crack openings along $N=2$ non-symmetrically located cracks in a cluster.}
\end{figure}

\begin{figure}[htp]
\centering
  \includegraphics[width=0.8\textwidth]{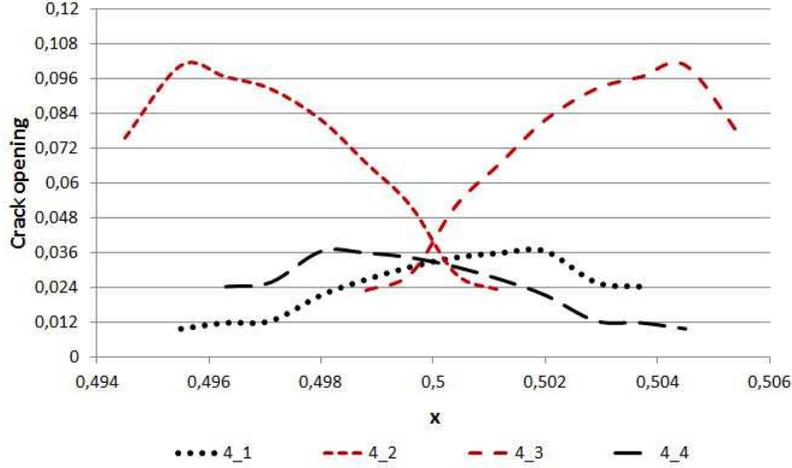}
  \caption{Crack openings along $N=4$ non-symmetrically located cracks in a cluster.}
\end{figure}

Nevertheless, in the case of non-symmetric cracks, the shadowing is weaker and consequently the results are more stable than in the case of symmetric configurations. The deterioration occurs when the ratio $d/L$ is notably smaller. Specifically, for $N=2$, $4$ and $6$ the results start to deteriorate when $d/L=0.0001$, $0.00075$ and $0.01$, respectively.

The distribution of the openings on a crack uniquely defines the SIFs at its tips. Hence, as clear from the results for openings, it can be expected that for the symmetrical configurations of Fig.1, the SIFs will be close to zero for the internal cracks, while for the external cracks they will be two-fold less than those for an isolated single crack of the same length. The results of Table 1 comply with this suggestion.

\begin{table}[tbh]
\caption{The values of the normalized SIF $K_{III}/K_{III}^0$ at the tips of $N=2$ and $N=4$ symmetrically located cracks, for decreasing distance $d$ between cracks.}\centering
\begin{tabular}[t]{|c||c|c||c|c|c|c|}
\hline
  & \multicolumn{2}{|c||}{$N=2$} & \multicolumn{4}{|c|}{$N=4$} \\ 
\cline{2-7}
$d/L$ & $2\_1$ & $2\_2$ & $4\_1$ & $4\_2$ & $4\_3$ & $4\_4$ \\ \hline
\cline{1-7}
0.67 & 0.8942 & 0.8942 & 0.7351 & 0.5949 & 0.5949 & 0.7351 \\ 
\cline{1-7}
0.33 & 0.8384 & 0.8384 & 0.6839 & 0.5357 & 0.5357 & 0.6839   \\
\cline{1-7}
0.06 & 0.7497 & 0.7497 & 0.5438 & 0.2646 & 0.2646 & 0.5438   \\ 
\cline{1-7}
0.03 & 0.6664 & 0.6664 & 0.5357 & 0.1372 & 0.1372 & 0.5357   \\ 
\cline{1-7}
0.006 & 0.5396 & 0.5396 & 0.5319 & 0.0077 & 0.0077 & 0.5319  \\ 
\cline{1-7}
0.003 & 0.5229 & 0.5229 & 0.5214 & 0.0015 & 0.0015 & 0.5214  \\
\cline{1-7}
\cline{1-7}
\textbf{single crack} & \multicolumn{2}{|c||}{1.014} & \multicolumn{4}{|c|}{1.014} \\ \hline
\end{tabular}
\end{table}
\medskip
They present the SIFs normalized by the value $K_{III}^0=\sigma_{yz}^0\sqrt{\pi l}$, corresponding to the SIF at the tips of a single crack with the half-length $l=L/2=0.075$, located at an infinite plate under unit anti-plane shear stresses $\sigma_{yz}^0=1.0$. (Note that we consider a finite square region, for which in the case of a single crack $K_{III}^S/K_{III}^0=1.014$). It can be seen that for the symmetrically located cracks, the normalized SIFs at the first and $N$-th crack tend to $K_{III}^S/2$, while the SIFs at each of the internal cracks tend to zero. Clearly the sum of the SIFs is close to the SIF of a single crack. Hence, the stresses at distances from the tips, exceeding the distance between the cracks, are practically the same as for a single crack located in the middle of a cluster. Still this equivalence does not mean that the conditions for the propagation of individual cracks are the same as well; they are quite different what has important implications discussed below. 

In the case of non-symmetrically located cracks, shown in Fig.1b, the SIFs at the tips, which are beyond the shadowed zones, tend to $K_{III}^0$. These are external tips of cracks $2\_1$ and $2\_2$ for the cluster of two cracks, and these are the external tips of the cracks $4\_2$ and $4\_3$ for the cluster of four cracks. The SIFs at the remaining (shadowed) tips tend to zero with decreasing distance between the cracks.

The results for the SIFs imply that at the stages of the hydraulic fracturing, for which the fracture toughness is significant, simultaneous symmetric propagation of a number of closely located parallel fractures is practically impossible for two reasons. Firstly, for such configurations, the driving SIFs are notably (at least two-fold) less than for a single crack. This makes symmetric propagation quite unfavorable. Secondly, if because of statistical fluctuations, one of the fractures overruns its closely located neighbors, it "shadows" the tips of the neighbors. Then SIFs at tips of the neighbors decrease to a level below the critical SIF, defined by the rock strength. Such an effect is especially significant at the stage of the fracture initiation what complies with the results for this stage obtained in the paper \cite{Zubkov2007}. 

This conclusion holds for the \textit{toughness dominated regime} of the fracture propagation. Meanwhile, for the \textit{viscosity dominated regime} we need to take into account for the changes  in the crack conductivity, rather than those in SIFs. 

In numerical simulations of hydraulic fractures, a fracturing fluid is commonly modelled as a Newtonian fluid. (When a fluid is non-Newtonian, it is assumed to be Newtonian with the dynamic viscosity $\mu$ defined by the characteristic strain rate; when a fluid contains proppant, the dynamic viscosity is prescribed from specially designed experiments as a function of the density of the slurry). Then the flux $q$ is connected with the pressure gradient $\partial p/ \partial x$ by the Poiseuille equation
$$
q=-k\frac{\partial p}{\partial x},
$$
where $k$ is the conductivity of a crack. It is essential that the conductivity strongly depends on the crack opening; for a single crack, it is proportional to the third degree of its opening $w_S$: 
$
k=k_{s}=w_S^3/(12\mu).
$
For a system of $N$ parallel cracks, the effective conductivity is the sum of individual conductivities:
\begin{equation}\label{equ8}
k=k_C=\sum_{i=1}^{N}w_{C_i}^3/(12\mu) ,
\end{equation}
where $w_{C_i}$ is the opening of the $i$-th fracture in a cluster. The resistance to a flow of a viscous fluid is inverse to $k$: $r=1/k$.

The results of the previous subsection serve us to evaluate the changes in the effective conductivity caused by the crack interaction. It is reasonable to compare the conductivity of a cluster $k_C$ with that of a single crack $k_{S}$ for various numbers $N$ of the cracks in a cluster and various normalized distances $d/L$ between the cracks:  
$$
\frac{k_C}{k_{S}}=\frac{\sum_{i=1}^N w_{C_i}^3}{w_S^3}.
$$

Fig.7 and Fig.8 present the results of calculations for cracks located in a cluster symmetrically and non-symmetrically, respectively. It can be seen that in all the cases the conductivity of a cluster is considerably less than that of a single crack. The ratio $k_C/k_{S}$ decreases as $1/(N^2)$. 

\begin{figure}[htp]
\centering
  \includegraphics[width=0.72\textwidth]{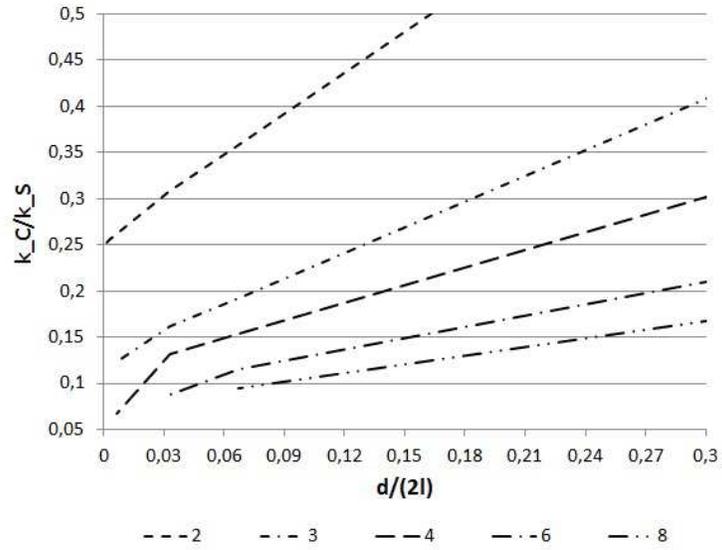}
  \caption{The dependence between the conductivity of multiple, symmetrically located cracks and a single crack, for various ratio $d/L$.}
\end{figure}
\begin{figure}[hbp]
\centering
  \includegraphics[width=0.72\textwidth]{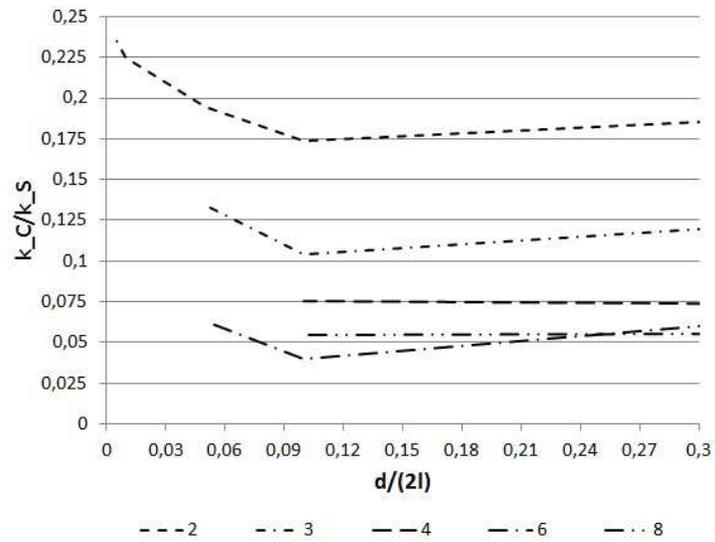}
  \caption{The dependence between the conductivity of multiple, non-symmetrically located cracks and a single crack, for various ratio $d/L$.}
\end{figure}

The decrease of the conductivity is the consequence of the fact that individual openings of closely located parallel cracks are notably less than the opening of a single crack. In view of the \textit{cubic} powers $w_{C_i}^3$ in (\ref{equ8}), this leads to fast decreasing of the effective conductivity. Then the total resistance of a cluster to viscous flow drastically increases as compared with the resistance of a single crack. As a result, the gradient of the pressure, needed for the transportation of a given volume of a fluid, is much less for a single crack of the width $w_S$ than the gradient needed to push the volume through a system of finer channels with the same sum $w_S$ of openings. Therefore localization of a prescribed flux in a single fracture appears to be much more preferable than its distribution in a number of finer channels with the same summary opening. 

We conclude that, for the \textit{viscosity dominated regime} of fracture propagation, the propagation of a number of closely located cracks is physically unfavorable. It implies that only those fractures, which are sufficiently far from other fractures of similar sizes may propagate. This conclusion agrees with the results of numerical simulations and field observations presented in the papers \cite{Damjanac2010}, \cite{Kresse2013}.

\section{Conclusions}

The conclusions of the study are summarized as follows. 

(i) Employing the CV H-BEM, with the second order approximation of the density functions and square root approximation of the opening near crack tips, may serve for studying closely located, nearly parallel cracks up to distances, at which the asymptotics for thin strips between the cracks become evident.

(ii) When the distance between cracks decreases or/and the number of cracks in a cluster grows, the crack interaction unavoidably leads to deterioration of numerical results. For the CV H-BEM used the results start to deteriorate when the condition number of the main matrix exceeds $10^{7}-10^{9}$.

(iii) The stress shadowing effect in a cluster of closely located nearly parallel cracks appears in actually uniform stresses and strains in thin strips between cracks. 

(iv) The openings of internal cracks in a cluster are nearly zero, while the openings of two external cracks are close to one-half of the opening of a single crack of the same size. Thus the sum of the openings in a cluster tends to the opening of a single crack when the number of internal cracks grows.
 
(v) The distribution of SIFs in a cluster is similar to that of the openings because the SIFs at tips of a crack are completely defined by the opening of the crack. Specifically, the SIFs of internal cracks are nearly zero, while the SIFs of two external cracks are close to one-half of the SIFs at the tips of a single crack of the same size. Thus the sum of the SIFs in a cluster tends to the SIF of a single crack when the number of internal cracks grows.

(vi) The fields of stresses, strains and displacements around a cluster of closely located nearly parallel cracks are close to those around a single crack located in the middle of the cluster. Consequently, the interaction between clusters may be modelled by placing merely one crack at the middle of each cluster. This approach may serve to avoid computational instability commented in the point (ii). 

(vii) The \textit{specific distribution of the SIFs} at tips of interacting cracks leads to propagation of a single fracture rather than to simultaneous propagation of a number of closely located fractures in the \textit{toughness dominated regime}. Similar effect occurs in the \textit{viscosity dominated regime} due to the \textit{specific distribution of the openings}, which results in drastic decrease of the conductivity of a cluster as compared with a single crack. 

(viii) Further investigation and an extension to clusters in 3D are needed to make quantitative recommendations for proper design of multiple hydraulic fractures in low-permeable shales.

\section*{Acknowledgement}
The authors gratefully acknowledge the support of the European Research Agency
(FP7-PEOPLE-2009-IAPP Marie Curie IAPP transfer of knowledge programme,
Project Reference \#251475).
\section*{References}


\bibliographystyle{model1a-num-names}
\bibliography{<your-bib-database>}

\end{document}